\documentclass{mem}

\usepackage{natbib}
\usepackage{txfonts}
\usepackage{balance}
\usepackage{graphicx}
\usepackage[a4paper]{hyperref}
\idline{00}{107}

\begin{document}

\title{Pattern speeds in interacting galaxies}

\subtitle{}

\author{C. L. \,Dobbs}

\offprints{Clare L. Dobbs; \email{dobbs@astro.ex.ac.uk}}
 
\institute{Department of Physics, University of Exeter, Exeter, UK}

\authorrunning{Dobbs}

\titlerunning{Interacting galaxies}

\abstract{We investigate pattern speeds in spiral galaxies 
where the structure is induced by an interaction with a companion
galaxy. We perform calculations modeling the response of the stellar
and/or gaseous components of a disc. Generally we do not find a unique
pattern speed in these simulations, rather the pattern speed decreases
with radius, and the pattern speed for individual spiral arms differ.
The maximum pattern speed is $\sim$20 km s$^{-1}$ kpc$^{-1}$ for the
discs with a live stellar component, decreasing to 5 km s$^{-1}$
kpc$^{-1}$ at the edge of the spiral perturbation. When only the gas
is modeled, $\Omega_{\rm p}$ is typically very low (5 km s$^{-1}$
kpc$^{-1}$) at all radii.
 }

\maketitle{}

\section{Introduction}

Most grand design spiral patterns are believed to be due to
interactions with other galaxies \citep{Toomre1972}, or are driven by
bars \citep[][Athanassoula et al., this volume]{Kormendy1979,
Bottema2003}. Here we investigate the first of these scenarios, by
performing numerical simulations of interacting galaxies. In this
instance, there is not necessarily a singular pattern speed for the
spiral arms, as indicated by recent observations of M51
\citep{Meidt2008a}.

\section{Method}

We use the Smoothed Particle Hydrodynamics Code (SPH) to model a
galaxy subject to an interaction. The first galaxy is modeled by
assuming a spherical potential for the halo, with 3 different
scenarios for the galactic disc: a) a live stellar disc, no gas; b) a
logarithmic stellar potential and a live gaseous disc and c) a disc
containing live gaseous and stellar components. The stars and/or gas
are allocated velocities such that the Toomre instability parameter,
Q, is globally 2, but in addition, the galaxy was allowed to evolve in
isolation until any flocculent structure disappears. In all
simulations 1 million particles are used. For the case of stars and
gas there are 500,000 gas particles and 500,000 stellar particles, and
in all cases the total mass of the disc is $5\times 10^9$ M$_{\odot}$.

We adopt a similar approach to \citet{Oh2008} to model the
interaction.  The interacting galaxy is represented by a sink particle
\citep{Bate1995} and is of relatively low mass, equal to the mass of
the disc or 2.5\% of the total mass of the first galaxy. The
interacting galaxy takes a parabolic orbit, reaching a closest
approach of 25 kpc after a time of 370 Myr. Initially this galaxy is
at a distance of 50 kpc.

\section{Results}
\subsection{Stellar disc}
\begin{figure}[]
\centering
\resizebox{\hsize}{!}{\includegraphics[clip=true]{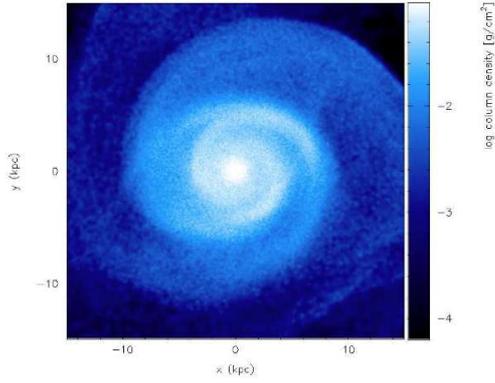}} 
\caption{\footnotesize
  The column density is shown for a stellar disc of a galaxy perturbed
  by a (point mass) galaxy on a parabolic orbit. The second galaxy is
  located 70 kpc away, at the coordinates $(-20,68)$ kpc. The stellar
  disc contains broad and diffuse spiral arms.}
\end{figure}

\begin{figure}[]
\centering
\resizebox{\hsize}{!}{\includegraphics[clip=true]{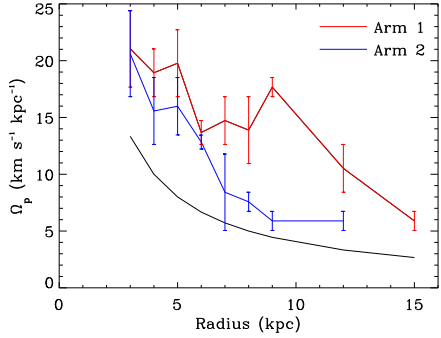}}
\caption{\footnotesize
  The pattern speeds for each spiral arm are plotted versus radius for
  the stellar disc. $\Omega_{\rm p}$ is different for each arm and decreases
  with radius. The lower line shows a $1/r$ dependence.}
\end{figure}

\begin{figure}[]
\centering
\resizebox{\hsize}{!}{\includegraphics[clip=true]{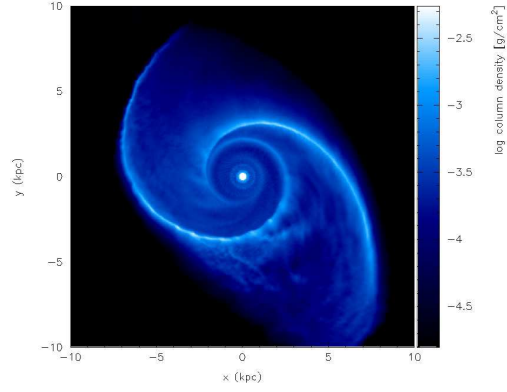}} 
\caption{\footnotesize
  The column density is shown for the gas disc perturbed by a galaxy,
  at the same time as Fig.~1. The gas shocks produce very narrow,
  dense spiral arms compared to Fig.~1.}
\end{figure}

\begin{figure}[]
\centering
\resizebox{\hsize}{!}{\includegraphics[clip=true]{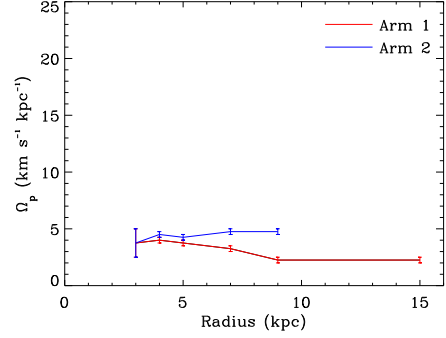}}
\caption{\footnotesize
  The pattern speeds are plotted for the spiral arms of the gas
  disc. $\Omega_p$ is much lower than typical from observations.}
\end{figure}

Fig.~1 shows the column density of a stellar disc, at a time of 820
Myr. The interacting galaxy is 70 kpc from the centre of the plot. The
spiral arms are relatively weak and broad.

We calculate the pattern speed of the spiral arms according to 
\begin{equation}
\Omega_{\rm p}=\frac{\phi(\rho_{\rm max})|_{t_2}-\phi(\rho_{\rm max})|_{t_1}}{t_2-t_1}
\end{equation}
where $\rho_{\rm max}$ is the peak density of a particular spiral arm
at a given radius, and $t_1$ and $t_2$ are times during the
simulation. First we select points covering a particular spiral arm at
time $t_1=800$ Myr, when a strong spiral pattern has emerged. Then the
azimuthal angle of the spiral arm is located for different radii. This
process is repeated at time $t_2=925$ Myr, to obtain the change in
azimuthal angle of the spiral arm at each radius, and thus
$\Omega_{\rm p}$. Given there are 2 spiral arms, this method leads to
a pattern speed for each spiral arm. These pattern speeds are shown
versus radius in Fig.~2, the errors reflecting the uncertainty in
locating the peak density of the spiral arm.

The magnitude of the pattern speeds are not dissimilar from some of
those measured for spirals (e.g., \citealt{Clemens2001, Grosbol2004}),
although the spiral arms clearly exhibit different pattern
speeds. This difference is a consequence of the asymmetry of the
system, i.e. that the interaction induces one arm on one side of the
disc first. The pattern speeds for each arm also decrease with radius,
roughly as expected for spiral patterns induced by interactions.

\subsection{Gaseous disc}

We also performed calculations with just gas. The gas constitutes 1\%
of the mass of the galactic disc. This value is unrealistically small,
but the low gas mass is chosen to avoid gravitational instabilities,
which would halt the calculation. Essentially, we are only
investigating the reaction of the gas to the interaction, not the self
gravity of the gas. These calculations are also not a particularly
realistic case as they ignore the perturbation experienced by the
stellar disc from the interaction (instead the stellar disc is
represented by a symmetric potential). However the case with just gas
is explored for completeness.

In Fig.~3, we show the disc when only the gas is included, at the same
time (800 Myr) as Fig.~1. The spiral arms are clearly much narrower,
more prominent, and more dense than for the stellar disc. The
evolution of the gas and stellar discs is also different. The gaseous
spiral arms rotate much slower than the stellar arms. Consequently the
pattern speed is very low, $\sim$4 km s$^{-1}$ kpc$^{-1}$ and does not
show much variation with radius (Fig.~4). The difference compared to
the stellar disc is that one spiral arm is still linked to, and
rotates at the same angular velocity as, the orbiting galaxy.

\begin{figure}[!h]
\centering
\includegraphics[width=.45\textwidth,clip=true]{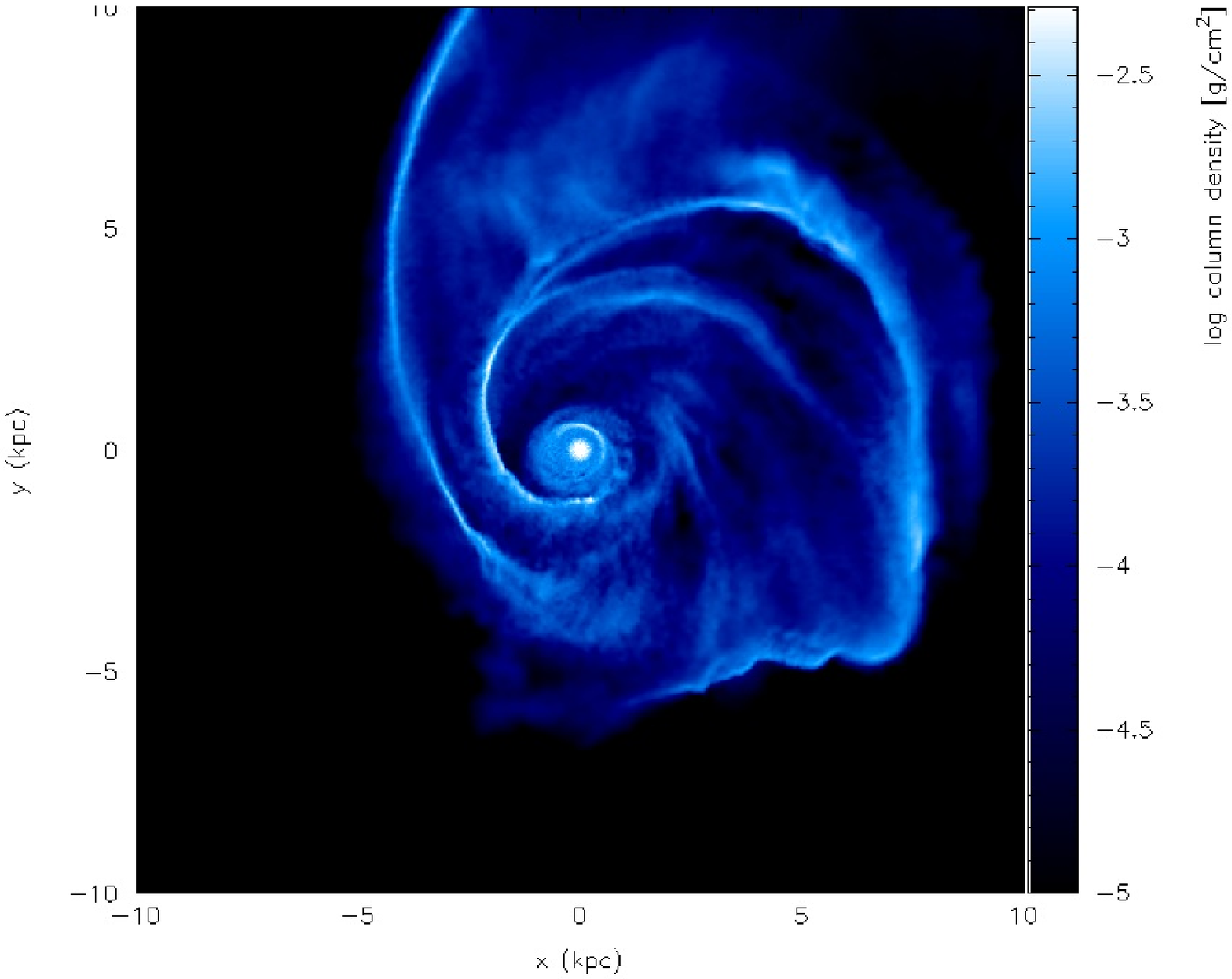} \\
\includegraphics[width=.45\textwidth,clip=true]{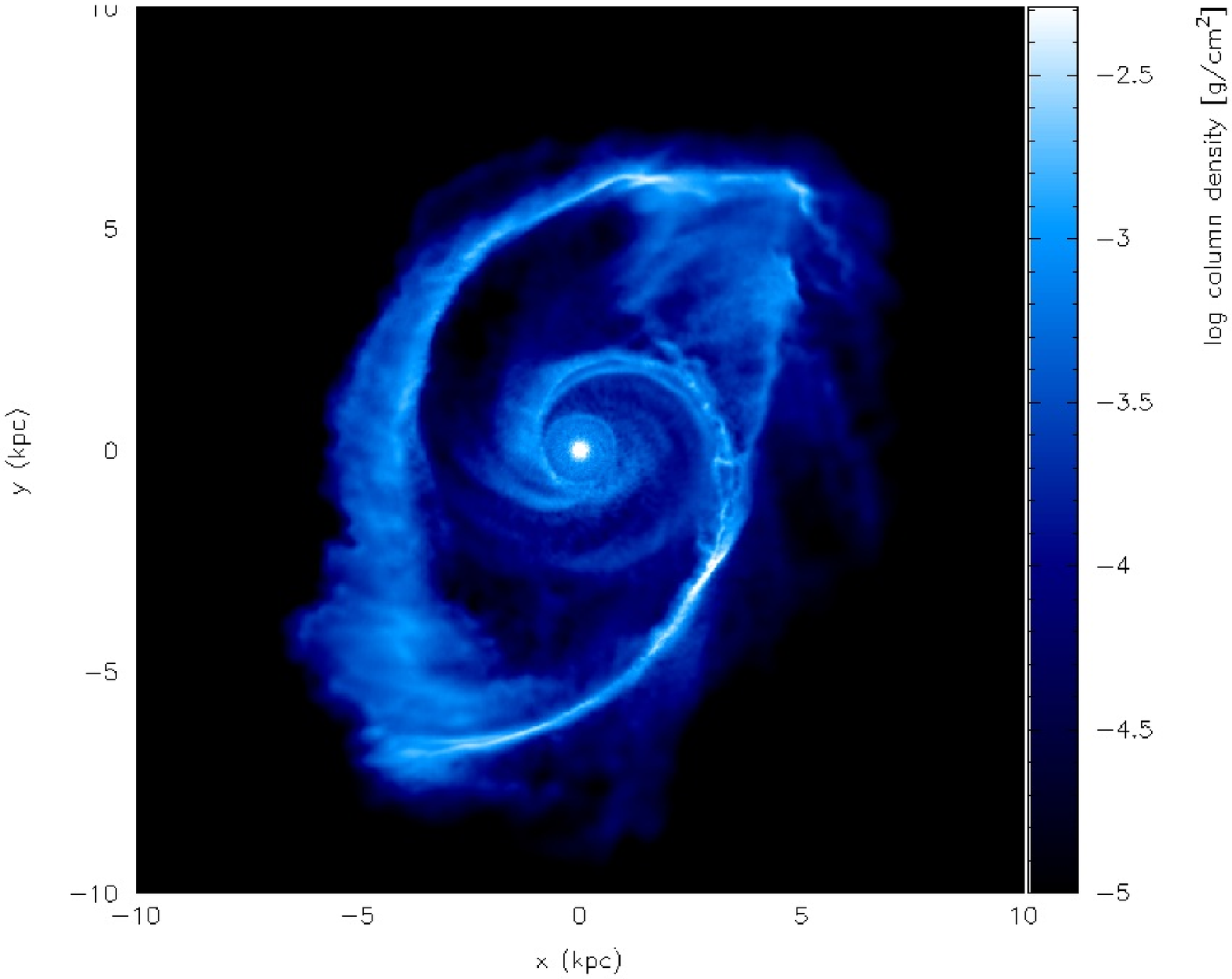} \\
\includegraphics[width=.45\textwidth,clip=true]{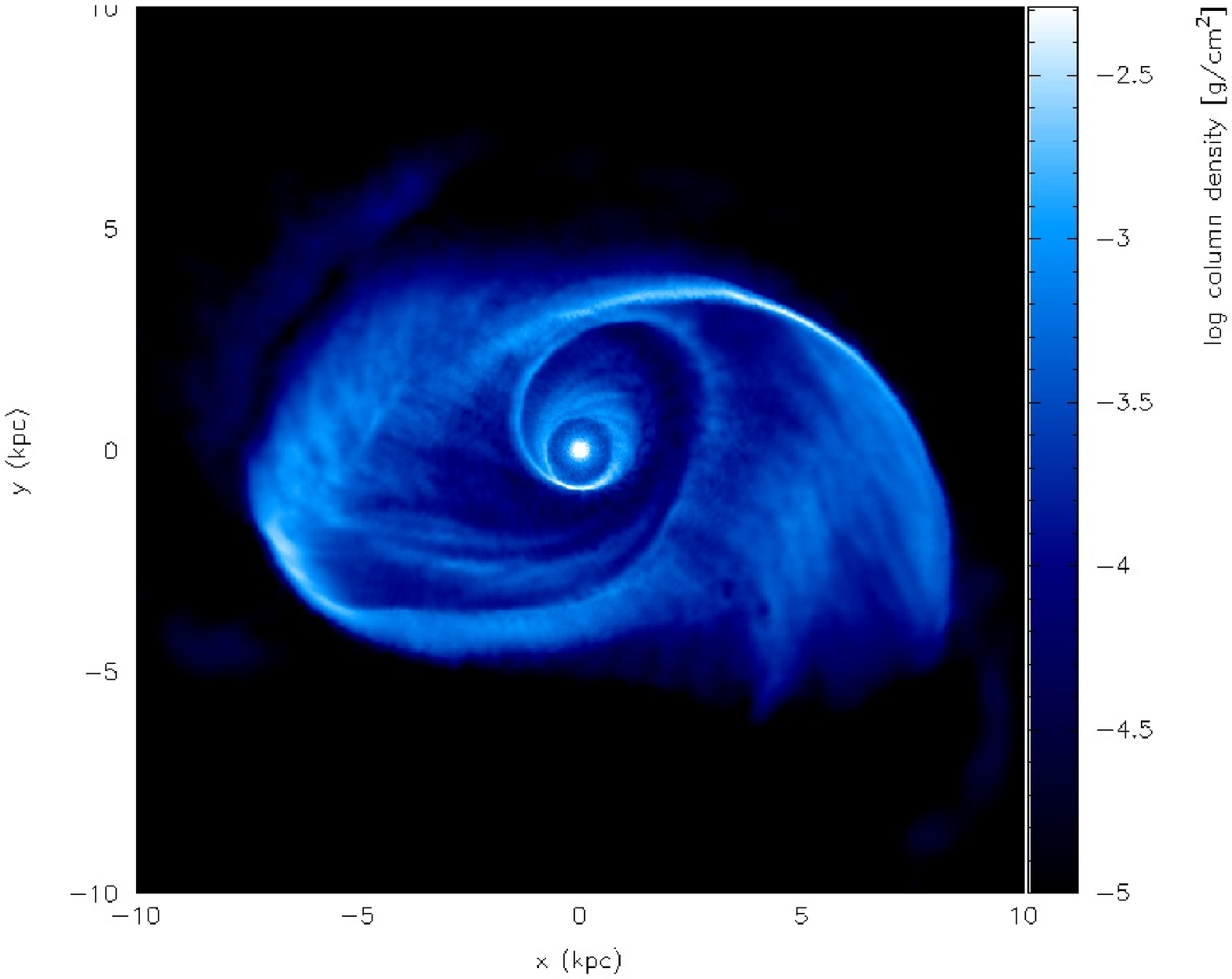}
\caption{\footnotesize
  The gas distribution is shown from a calculation with gas and stars
  at times of 500 (top), 740 (middle) and 1060 Myr (bottom). The
  spiral arms are very asymmetric at the earliest time (when the
  position of the interacting galaxy is located at $(22.5, 15)$ kpc),
  but becomes more symmetric at later times, after the interaction.}
\end{figure}

\begin{figure}[]
\centering
\resizebox{\hsize}{!}{\includegraphics[clip=true]{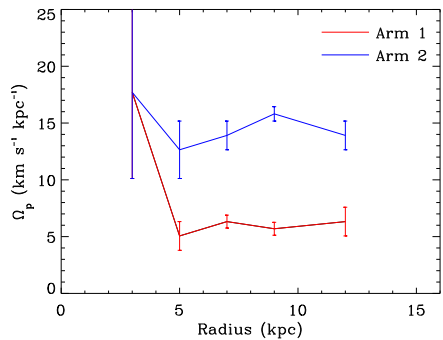}} 
\caption{\footnotesize
  The pattern speed is shown for the gaseous arms, from the simulation
  with stars and gas. The gas is largely coupled to the stars, hence
  $\Omega_{\rm p}$ is higher than when only gas particles are
  present. Again there is a clear difference in the pattern speeds of
  each spiral arm.}
\label{li_vhel}
\end{figure}

\subsection{Evolving both the stars and gas}

Finally we show calculations with gas and stars, again where the gas
represents 1\% of the mass of the stellar disc. Thus the mass of the
stellar disc is $5\times 10^9$ M$_{\odot}$ and the gaseous disc $5
\times 10^8$ M$_{\odot}$, although the actual number of gas and
stellar particles in the calculation are equal.  Fig.~5 shows the
column density of gas from the calculation with live stellar and
gaseous components.  The morphology evidently changes over time. At an
earlier stage in the interaction, the spiral pattern is very
asymmetric (unlike the calculations with gas alone, where the spiral
pattern is symmetric throughout the simulation). The pattern becomes
more symmetric at later times, tending towards the distribution
without a live stellar disc (Fig.~3). Although not shown on Fig.~5,
the stellar distribution reflects the gas distribution, but the spiral
arms are much weaker and broader.

Finally in Fig.~6 we plot the pattern speed of the gas from the
calculation with both live stellar and gaseous components. The times
selected to calculate the pattern speed are 450 and 500 Myr. The
spiral arms clearly have very different pattern speeds, which is not
surprising given the asymmetry of the disc. The spiral arm with the
lower pattern speed is still associated with the interacting
galaxy. The pattern speeds are more similar in magnitude to the case
when only stars are used (Fig.~2), since the gas is better coupled to
the stars.

\section{Conclusions}

We have performed calculations with a stellar, gaseous and both
stellar and gaseous disc subject to an interaction with an orbiting
galaxy. The pattern speed across the disc is generally not constant,
and pattern speeds in each arm differ. For a stellar disc,
$\Omega_{\rm p}=5-20$ km s$^{-1}$ kpc$^{-1}$, decreasing with radius
approximately as $1/r$. With only gas, the pattern speeds are much
lower ($3-6$ km s$^{-1}$ kpc$^{-1}$). When stars and gas are included,
the gas tends to follow the stellar distribution, thus the pattern
speeds of the gaseous spiral arms are higher ($5-17$ km s$^{-1}$
kpc$^{-1}$).

These calculations may be improved by using a more consistent initial
galaxy set up (e.g., \citealt{Dubinski1995}). A natural extension of
this work would also be to compare with observations by applying the
Tremaine-Weinberg \citep{Tremaine1984} method to these calculations
(see also \citealt{Meidt2008b}).
 
\begin{acknowledgements}
The calculations presented in these proceedings were performed using
the University of Exeter's SGI Altix ICE 8200 supercomputer, Zen.
\end{acknowledgements}


\bibliographystyle{aa}

\end{document}